\documentclass[twocolumn,showpacs,preprintnumbers,amsmath,amssymb,pra,superscriptaddress]{revtex4-1}
\usepackage{amsfonts}
\usepackage[english]{babel}
\usepackage[T1]{fontenc}
\usepackage{times}
\usepackage{mathrsfs}
\usepackage{graphicx}
\usepackage{dcolumn}
\usepackage{bm}
\usepackage{makecell}
\usepackage[colorlinks,bookmarks=true,citecolor=blue,linkcolor=red,urlcolor=blue]{hyperref}
\usepackage[tight, FIGTOPCAP, hang, raggedright, nooneline]{subfigure}
\begin{document}


\title{Fermionic non-Abelian fractional Chern insulators from dipolar interactions}
\author{Dong Wang}
\affiliation{Institute of Physics, Chinese Academy of Sciences, Beijing 100190, China}
\author{Zhao Liu}
\email{zhaol@princeton.edu}
\affiliation{Department of Electrical Engineering, Princeton University, Princeton, New Jersey 08544, USA}
\author{Wu-Ming Liu}
\affiliation{Institute of Physics, Chinese Academy of Sciences, Beijing 100190, China}
\author{Junpeng Cao}
\affiliation{Institute of Physics, Chinese Academy of Sciences, Beijing 100190, China}
\affiliation{Collaborative Innovation Center of Quantum Matter, Beijing 100190, China}
\author{Heng Fan}
\email{hfan@iphy.ac.cn}
\affiliation{Institute of Physics, Chinese Academy of Sciences, Beijing 100190, China}
\affiliation{Collaborative Innovation Center of Quantum Matter, Beijing 100190, China}

\date\today

\begin{abstract}
We study fermions on a triangular lattice model that exhibits topological flatbands characterized by nonzero Chern numbers. Our scheme stems from the well-known Hofstadter model but the next-nearest-neighbor hopping is introduced, which is crucial for tuning the lowest band to be nearly flat. Differing from previous proposals with the necessity of multi-particle interactions, we consider the more realistic long-range dipolar interaction combined with two-body short-range attractions between fermions. We show the realization of the non-Abelian $\nu=1/2$ Moore-Read fractional Chern insulators, and strong evidence for the existence of the more exotic $\nu=3/5$ Read-Rezayi fractional Chern insulators. Our results provide insights for the experimental realization of these exotic states by realistic two-body interactions and thus facilitates the implementation of the universal topological quantum computation.
\end{abstract}

\maketitle

\section{Introduction}
The cousins of fractional quantum Hall (FQH) effect \cite{PhysRevLett.50.1395,PhysRevLett.48.1559} on two-dimensional (2D) lattices have been attracting great interest recently. In a Chern band possessing a nonzero Chern number as an analog of a single Landau level \cite{PhysRevLett.106.236802,PhysRevLett.106.236803,PhysRevLett.106.236804}, the interaction of particles in the fractionally filled band leads to strongly correlated states named fractional Chern insulators (FCIs) \cite{review1,review2,review3,sheng2011fractional,PhysRevX.1.021014,liu2012fractional,yao2013realizing, PhysRevLett.110.185301,wu2013bloch,andreasprl,wang2013tunable}. Compared with their FQH counterparts, a strong net external magnetic field is no longer an indispensable element, and FCIs are expected to be much more robust against high temperature \cite{PhysRevLett.106.236802}.

Among various FCIs, the most exotic members are those that support excitations, i.e., anyons, obeying non-Abelian statistics \cite{Moore1991362,PhysRevLett.66.802}. These non-Abelian anyons are essential resources in topological quantum computation \cite{Nayak}. However, the realization of non-Abelian FCIs (as well as non-Abelian FQH states) in realistic models is usually very difficult. Up to now, except in a few cases of bosons \cite{PhysRevLett.110.185301,PhysRevB.88.081106,PhysRevB.88.205101}, almost all numerically confirmed non-Abelian FCIs are stabilized by peculiar multi-particle interactions \cite{PhysRevLett.108.126805,Bernevig_PRB85,zoology,PhysRevB.87.205137,Weyl,greiter2009non,greiter2014parent} (the stabilization of bosonic non-Abelian FQH states is also usually much easier than that of the corresponding fermionic states). Considering that electronic materials and fermions in optical lattices are natural platforms for Chern bands, the discovery of non-Abelian FCIs stabilized by realistic two-body interactions in fermionic systems is highly demanded. This construction is of fundamental interest and facilitates the future implementation of topological quantum computation.

In this paper, we report progress in this direction. We choose a simple generalization of the triangular Hofstadter model. By introducing the next-nearest neighbor hopping, the lowest Bloch band can be tuned to be very flat even for a large value of flux density. This is a compelling feature that makes this model an ideal platform to search for non-Abelian FCIs. Enlightened by the positive effect of long-range interactions on the stabilization of bosonic non-Abelian FCIs \cite{PhysRevB.88.205101}, we turn on the experimentally realistic dipolar interaction \cite{DipolarExp1,DipolarExp2,DipolarExp3} between fermions, supplemented by two-body short-range attractions that might be controlled by Feshbach
resonances \cite{PhysRevLett.103.080406}. By using exact diagonalization, we study the many-body system in several aspects, such as the energy spectrum, the particle-cut entanglement spectrum \cite{PhysRevLett.101.010504,PhysRevLett.106.100405,PhysRevX.1.021014}, and the adiabatic continuity to the system with multi-particle interactions.
We obtain convincing numerical results to confirm the existence of the non-Abelian $\nu=1/2$ Moore-Read FCIs \cite{Moore1991362}. Through the analysis of two-particle energy spectrum, we show that our choice of the attraction strength is reasonable to stabilize the Moore-Read FCIs. Some encouraging evidence that supports the \emph{Z}$_3$ $\nu=3/5$ Read-Rezayi FCIs \cite{PhysRevB.59.8084} is also discovered. The stabilization of the fermionic $\nu=3/5$ Read-Rezayi FCI is very exciting because its Fibonacci anyon excitation is necessary for universal quantum computation.

\section{Single-particle model and band topology} 
\begin{figure}
\centerline{\includegraphics[width=1.0\linewidth] {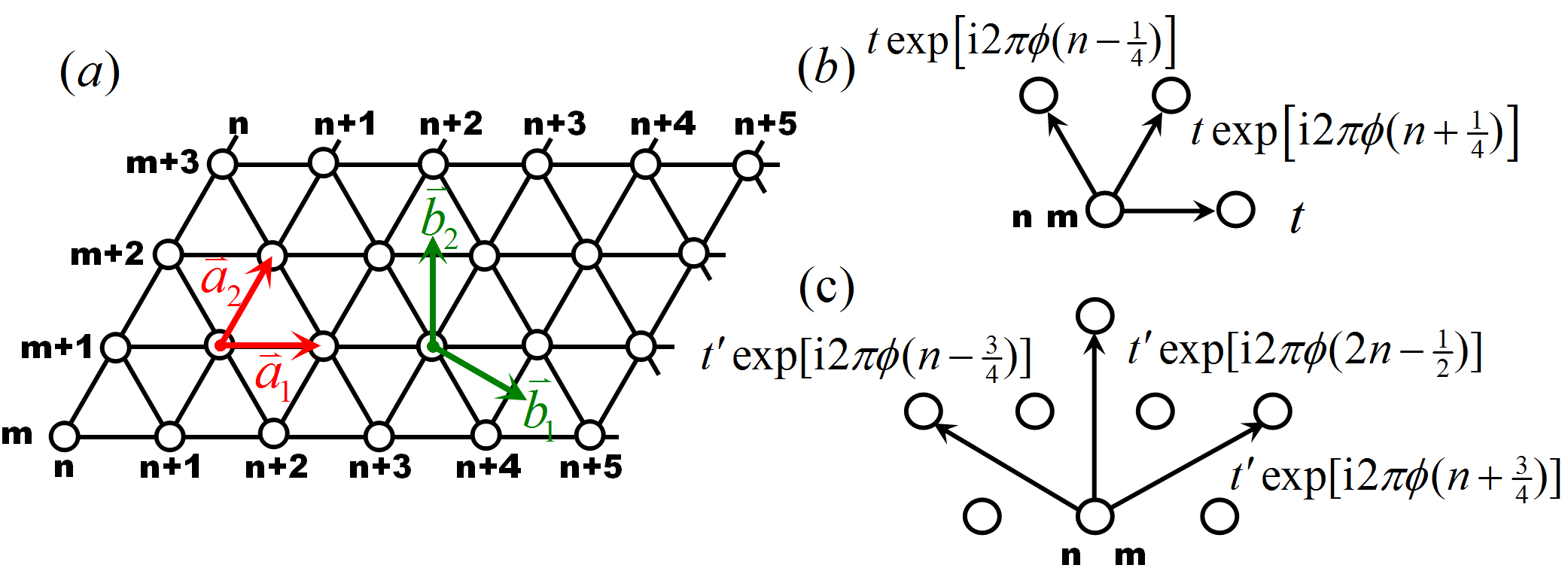}}
\caption{(Color online) (a) The schematic graph of our triangular lattice model with $(\vec a_1,\vec a_2)$ and $(\vec b_1,\vec b_2)$ representing lattice vectors and reciprocal lattice vectors respectively. $(m,n)$ labels the lattice site. (b) Complex hopping amplitudes between NN sites. (c) Complex hopping amplitudes between NNN sites. One can easily verify that the magnetic flux per each plaquette (two triangles) is $\phi$ in units of the flux quantum $\phi_0$.}
\label{fg:lattice}
\end{figure}

We consider spinless fermions on a 2D triangular lattice penetrated by an uniform magnetic field. Assuming fermions only hop between nearest-neighbor (NN) and next-nearest-neighbor (NNN) sites (Fig.~\ref{fg:lattice}), the single-particle Hamiltonian is
\begin{equation}
\label{eq:H_0}
H_0=-\sum_{\langle i,j\rangle,\langle\langle i,j \rangle\rangle}t_{ij}e^{\textrm i\phi_{ij}}c^\dagger_i c_j,
\end{equation}
where $c_i$ ($c_i^\dagger$) is the fermionic annihilation (creation) operator on site $i$, $t_{ij}=t$ for NN sites and $t_{ij}=t'$ for NNN sites, and $\phi_{ij}$ is indicated in Fig.~\ref{fg:lattice}.
Our model is actually the triangular version of the well-known Hofstadter model \cite{hof} with extra hopping between NNN sites.

If the magnetic flux density $\phi=p/q$, where $p$ and $q$ are coprime integers, each unit cell consists of $q$ sites in the $\vec a_1$ direction. The band structure should exhibit $q$ Bloch bands, each of which can be labeled by a Chern number \cite{tknn}. For the special case of $p=1$, all bands are separated from each other by finite gaps. Their Chern numbers can be described by a simple picture: the $(q-1)$ lower bands have unit Chern number $C_{i<q}=1$, while the $q$th (highest) band has Chern number $C_q=-q+1$, satisfying $\sum_{i=1}^qC_i=0$. This Chern number distribution is different from the one on the square lattice \cite{mk,wang2013tunable} because of the different butterfly structures of the energy spectra.

\begin{figure}
\centerline{\includegraphics[width=\linewidth]{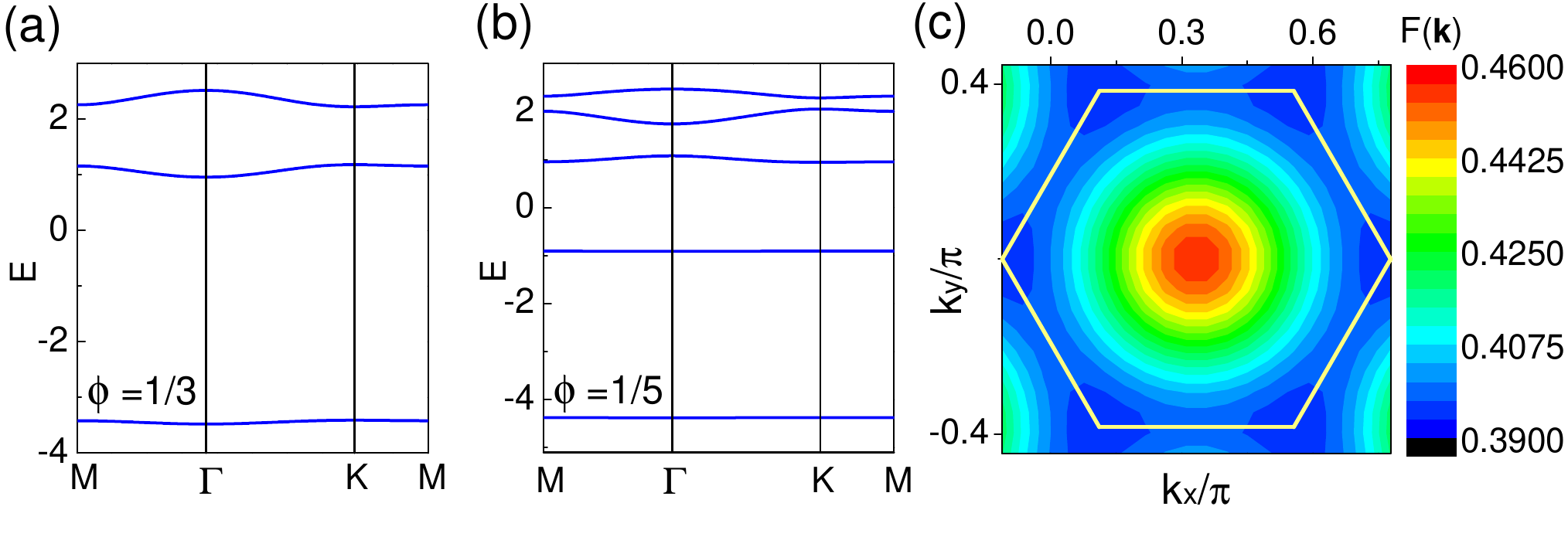}}
\caption{(Color online) Band structure for (a) $H_0(\phi{=}\frac13)$, $t'/t=0.16$; and (b) $H_0(\phi{=}\frac15)$, $t'/t=0.10$. (c) The Berry curvature of the lowest band for $H_0(\phi{=}\frac13)$ with $t'/t=0.16$. The region within the white solid line is one third of a Brillouin zone.
\label{fg:onebody}}
\end{figure}

By optimizing the ratio of $t'/t$, we can tune the lowest band to be nearly flat. For example, when $\phi=1/3$ and $t'/t=0.16$, the flatness of the lowest band, defined as the ratio between the band gap $\Delta$ and the bandwidth $w$, reaches $\frac{\Delta}{w}\approx64$ [Fig.~\ref{fg:onebody}(a)]. With larger $q$, we can even obtain more than one flatband. For $\phi=1/5$ and $t'/t=0.10$, the flatness of the lowest two bands are $\frac{\Delta_1}{w_1} \approx 1449$ and $\frac{\Delta_2}{w_2} \approx 284$ respectively [Fig.~\ref{fg:onebody}(b)]. Interestingly, the fluctuation of the Berry curvature $F$ of the lowest band is much smaller than that on the square lattice with the same flux density \cite{PhysRevB.88.205101}. For $\phi=1/3$ and $t'/t=0.16$, $F$ of the lowest band is quite uniform [Fig.~\ref{fg:onebody}(c)] and close to the mean value $\overline F= \frac{\sqrt{3}q}{4\pi} \approx 0.4135$. This suggests that the lowest band of our triangular lattice model is a more suitable host for fractional Chern insulators than that of the square lattice model.

\section{Moore-Read FCIs at $\nu=1/2$}
Now we consider $N_{e}$ interacting fermions partially filled in the lowest flatband on the torus. We assume dipolar potential $v(\mathbf r)=1/|\mathbf r|^3$ between fermions, which is experimentally realistic for neutral fermions and can be realized by trapping ultracold polar molecules in optical lattices \cite{DipolarExp1,DipolarExp2,DipolarExp3}. We also include short-range two-body attractive Hubbard terms between $n$-th NN terms [$n=1$ corresponds to the NN interaction, $n=2$ corresponds to the NNN interaction, etc.]. It has been proposed in Ref.~\cite{PhysRevLett.103.080406} that the $n=1$ NN term due to the $s$-wave scattering between fermions in optical lattices can be controlled by Feshbach
resonances. The whole two-body interaction Hamiltonian is
\begin{equation}\label{eq:H_LR}
H_{2\textrm{b}}=\sum_{i<j}V_{\textrm{d-d}}(\mathbf r_i-\mathbf r_j)n_in_j-\sum_{m=1}^{n_{\textrm{max}}} U_m \Big(\sum_{(i,j)\in\mathcal N_m}n_in_j\Big),
\end{equation}
where
\begin{equation}
V_{\textrm{d-d}}(\mathbf r)=\!\!\!\sum^{+\infty}_{s,r=-\infty}\!\!v(\mathbf r+sN_1\vec a_1+rN_2\vec a_2)
\end{equation}
that is periodic for $N_1\times N_2$ lattice sites.
We assume that the strength of interaction is much smaller than the band gap but larger than the bandwith, so $H_0$ is quenched and we can project $H_{2\textrm{b}}$ onto the occupied lowest band. We diagonalize the projected Hamiltonian $H_{2\textrm{b}}$. Because each magnetic unit-cell contains $q$ sites, there are $\emph{\b{N}}_1\times\emph{\b{N}}_2$ unit cells with $\emph{\b{N}}_{1}=N_1/q$ and $\emph{\b{N}}_{2}=N_2$. The band filling factor $\nu$ is defined as $\nu= {N_e} /(\emph{\b{N}}_{1}\emph{\b{N}}_{2})$. Since the total translation operator commutes with both $H_0$ and $H_{2\textrm{b}}$, each energy level can be labeled by a 2D total momentum $(K_1,K_2)$ with $K_{1,2}=0\thicksim {(\emph{\b{N}}_{1,2} {-} 1)}$.

\begin{figure}
\centerline{\includegraphics[width=1.0\linewidth]{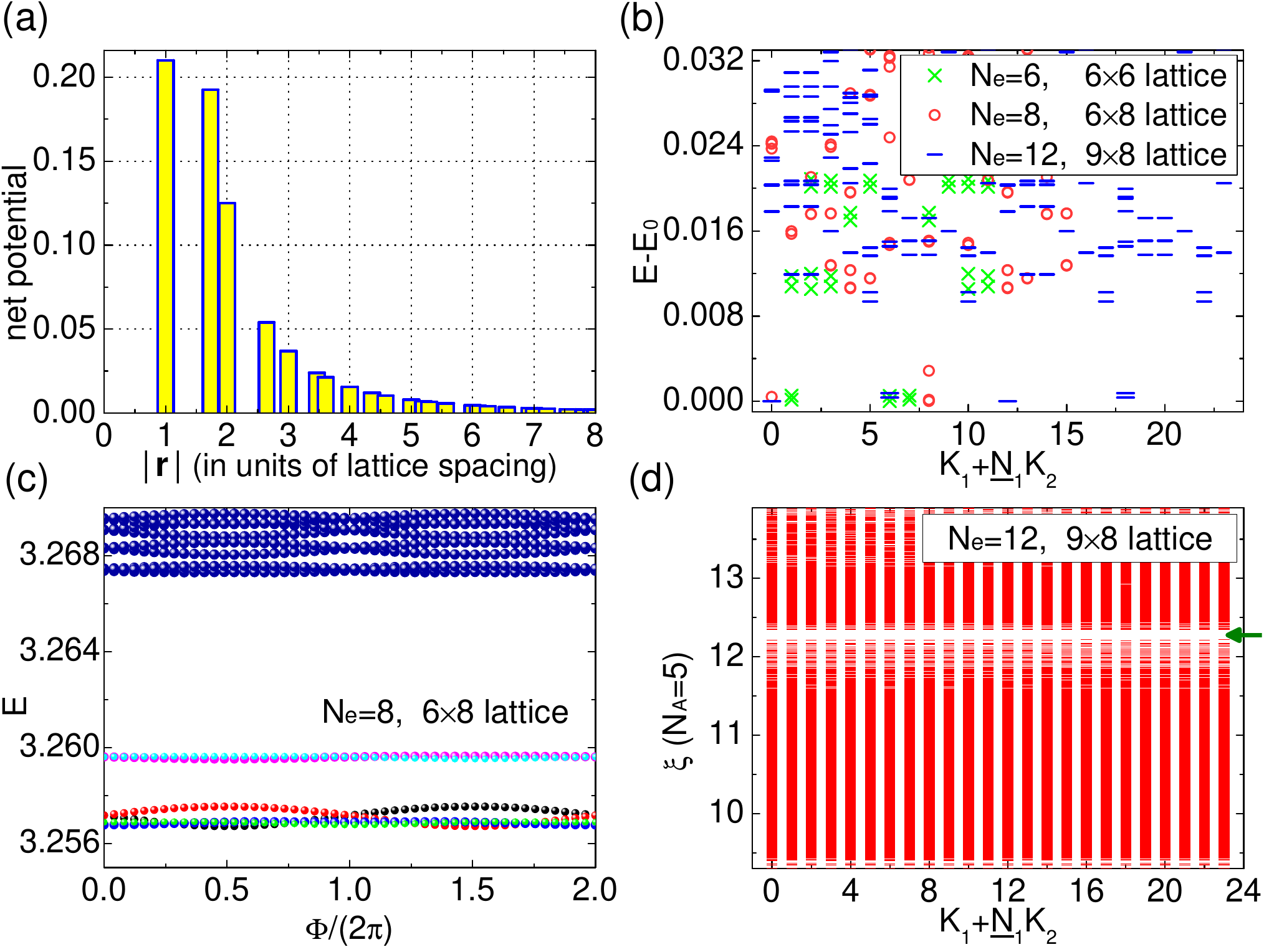}}
\caption{(Color online) Evidence of the $\nu=1/2$ MR FCIs as the ground states of $H_{2\textrm b}$ with $\phi=1/3$, $t'/t=0.16$, $n_{\textrm{max}}=1$, and $U_1=0.79$. (a) The net potential for the combination of dipolar interaction and two-body attractive NN interaction as a function of distance $r$ on the lattice. (b) The low-energy spectra at $\nu=1/2$ for $N_{e}=6,8,12$ on $N_1\times N_2=6\times6$, $6\times8$, $9\times8$ lattices, respectively. (c) The $x$-direction spectral flow for $N_{e}=8$ on the $N_1\times N_2=6\times8$ lattice. (d) The PES for $N_{e}=12$ and $N_{A}=5$ on the $N_1\times N_2=9\times8$ lattice. The number of states below the gap (indicated by the green arrow) is 30648.
\label{fg:LR_halffilling}}
\end{figure}

We first focus on $\nu{=}1/2$ to look for the non-Abelian Moore-Read (MR) FCIs. Compelling evidence, as displayed in Fig.~\ref{fg:LR_halffilling}, demonstrates that the ground states are indeed in the MR phase for flux densities as high as $\phi=1/3$. By choosing $n_{\textrm{max}}=1$, $U_1=0.79$ to weaken the repulsion between NN sites [Fig.~\ref{fg:LR_halffilling}(a)], we observe six quasidegenerate ground states for each system size that we study, and they are separated from the excited levels by an energy gap much larger than the ground-state splitting [Fig.~\ref{fg:LR_halffilling}(b)]. This degeneracy is robust against the twisted boundary conditions, i.e., the six ground states never mix with excited levels in the spectral flow [Fig.~\ref{fg:LR_halffilling}(c)]. In order to further investigate the topological order of the ground-states, we compute the commonly used particle-cut entanglement spectrum (PES) \cite{PhysRevLett.106.100405,PhysRevX.1.021014,Bernevig_PRB85} to rule out the possibility of other effects such as the charge density wave.
After dividing the whole system into two parts $A$ and $B$ with $N_{A}$ and $N_{B}$ particles respectively, and tracing out part $B$ from the density matrix $\rho=\frac1m \sum^m_{i{=}1}|\Psi^i\rangle\langle\Psi^i|$ of the ground state manifold, where $|\Psi^i\rangle$ represents the $i$-th state of $m$ degenerate ground states in the manifold, we can obtain the PES level defined as $\xi_i = -\ln\lambda_i$ with $\lambda_i$ the eigenvalues of the reduced density matrix $\rho_{ A}=\textrm{Tr}_{B}\rho$.
We find that a clear gap (which may increase for smaller $N_A$) exists in the PES, below which the number of levels matches the quasihole excitation counting of the MR state predicted by the $(2,4)-$admissible rule \cite{Bernevig_PRB85} [Fig.~\ref{fg:LR_halffilling}(d)]. All these results above conclusively confirm the existence of the $\nu=1/2$ MR FCIs in the presence of dipolar interaction and attractive NN interaction (\ref{eq:H_LR}) at high flux densities.

Similar results can also be obtained for different system sizes and flux densities. For example, we can choose a smaller flux density $\phi=1/5$ and project $H_{2\textrm{b}}$ to either the lowest or the second lowest flatband [Fig.~\ref{fg:onebody}(b)]. As both of these two flatbands have unit Chern number, we can stabilize the $\nu=1/2$ MR FCIs on each of them. Which band is fractionally filled is determined by the chemical potential.

\begin{figure}
\centerline{\includegraphics[width=1.0\linewidth]{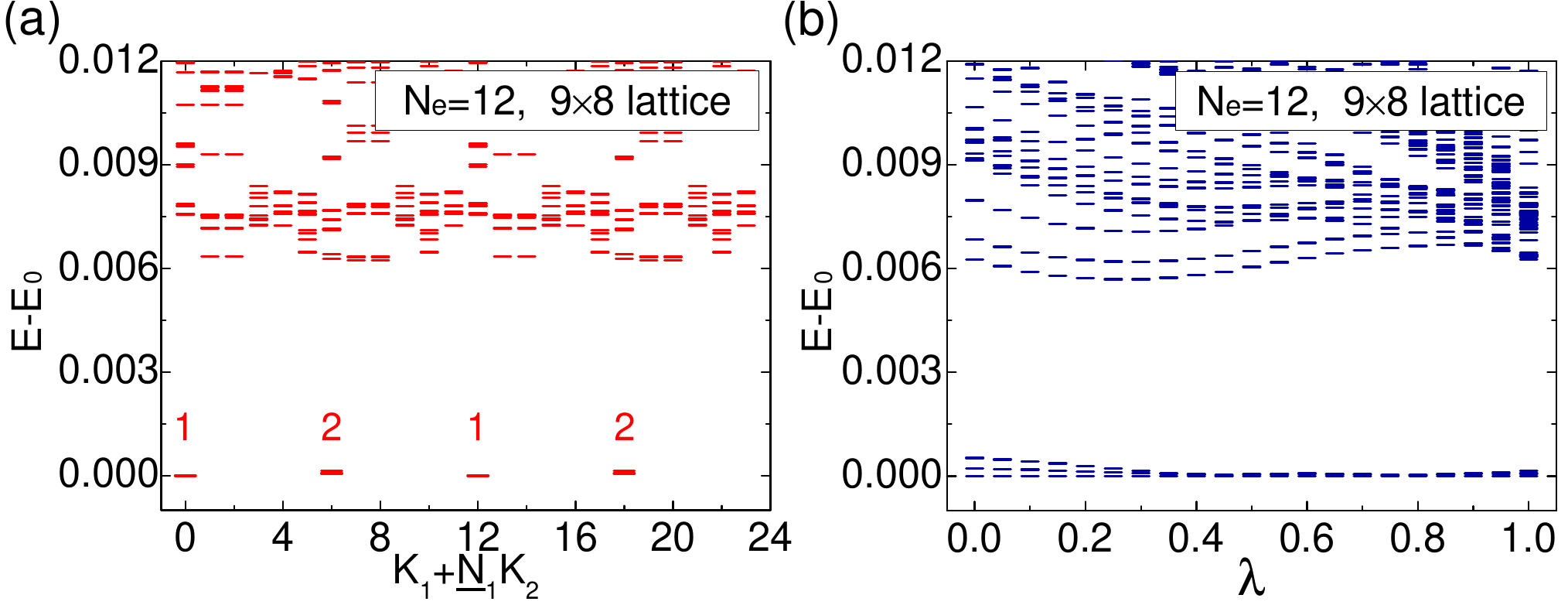}}
\caption{(Color online) The adiabatic continuity between the ground states of $H_{3\textrm b}$ and $H_{2\textrm b}$. We choose $U_{3\textrm b}=1.16$ in $H_{3\textrm b}$ and $\phi=1/3$, $t'/t=0.16$, $n_{\textrm{max}}=1$, $U_1=0.79$ in $H_{2\textrm b}$. (a) The low-energy spectrum of $H_{\textrm{3b}}$ for $N_{ e}=12$ on the $N_1\times N_2=9\times8$ lattice. The number of ground states in each sector is labeled on the chart. (b) The evolution of the low-energy spectrum of $H_{\textrm{int}}(\lambda)$ with $\lambda$ from $0$ to $1$ for $N_{e}=12$ on the $N_1\times N_2=9\times8$ lattice. The energy gap is made equal at $\lambda=0$ and $\lambda=1$.\label{fg:continuty}}
\end{figure}

\section{Adiabatic continuity to the ground states of three-body interactions}
It is already known that the $\nu=1/2$ MR FCIs can be stabilized by NN three-body repulsive interactions \cite{zoology,Bernevig_PRB85}. In our lattice model, we consider the three-body repulsion on each triangular plaquette, i.e.,
\begin{equation}
\label{3bhamil}
H_{3\textrm{b}}\!=U_{3\textrm{b}}\sum_{\langle i,j,k\rangle\in\bigtriangleup,\bigtriangledown} n_i n_j n_k
\end{equation}
with $U_{3\textrm{b}}>0$. The low-energy spectrum of this three-body interaction for $N_{ e}=12$ fermions at $\nu=1/2$ with $\phi=1/3$ is displayed in Fig.~\ref{fg:continuty}(a). As expected, we find a ground-state manifold of six-fold degeneracy. Moreover, further analysis including the quasihole excitations and PES supports that the ground states are indeed in the MR phase.

The existence of MR FCIs for the three-body interaction enables us to investigate the adiabatic continuity between the ground states of $H_{2\textrm{b}}$ and $H_{3\textrm{b}}$. In order to achieve this, we construct a Hamiltonian interpolating between $H_{2\textrm{b}}$ and $H_{3\textrm{b}}$, i.e.,
\begin{equation}
H_{\textrm{int}}(\lambda)=(1-\lambda)H_{2\textrm{b}}+\lambda H_{3\textrm{b}},
\end{equation}
where $\lambda\in[0,1]$ is the interpolation parameter. When $\lambda$ continuously increases from $0$ to $1$, this Hamiltonian evolves from $H_{2\textrm{b}}$ to $H_{3\textrm{b}}$. We diagonalize $H_{\textrm{int}}(\lambda)$ to study how does the energy spectrum evolve with $\lambda$. The result for $N_{e}=12$ fermions is shown in Fig.~\ref{fg:continuty}(b). The parameters in $H_0$ and $H_{2\textrm b}$ are the same as those used in the previous section. During the interpolation from $\lambda=0$ to $\lambda=1$, we find that there are always six quasidegenerate ground states well separated from high excited levels. The fact that the energy gap does not close during the interpolation suggests that the ground states at $\lambda=0$ and $\lambda=1$ are adiabatically connected and in the same phase. This adiabatic continuity provides another convincing evidence that the ground states of the two-body long-range interaction $H_{2\textrm{b}}$ at $\nu=1/2$ are indeed MR FCIs.

\section{Two-particle spectrum analysis}
In the Landau level physics, any rotation (translation) invariant two-body Hamiltonian is determined by its Haldane's pseudopotential parameters $\mathcal{V}_m$ \cite{pseudopotential}, which can be calculated analytically. Although we do not have an elegant formula for Chern band like for Landau levels, we can still approximately extract pseudopotential parameters from the energy spectrum of two interacting particles \cite{andreasprl}. These pseudopotential parameters in the Chern band provide guidance for what interaction we should use to stabilize a target FCI \cite{PhysRevB.88.205101}. Therefore, in order to further understand why the Hamiltonian $H_{2\textrm{b}}$ can stabilize the $\nu=1/2$ MR FCIs, we consider the two-fermion problem.

\begin{figure}
\centerline{
\includegraphics[width=1.0\linewidth]{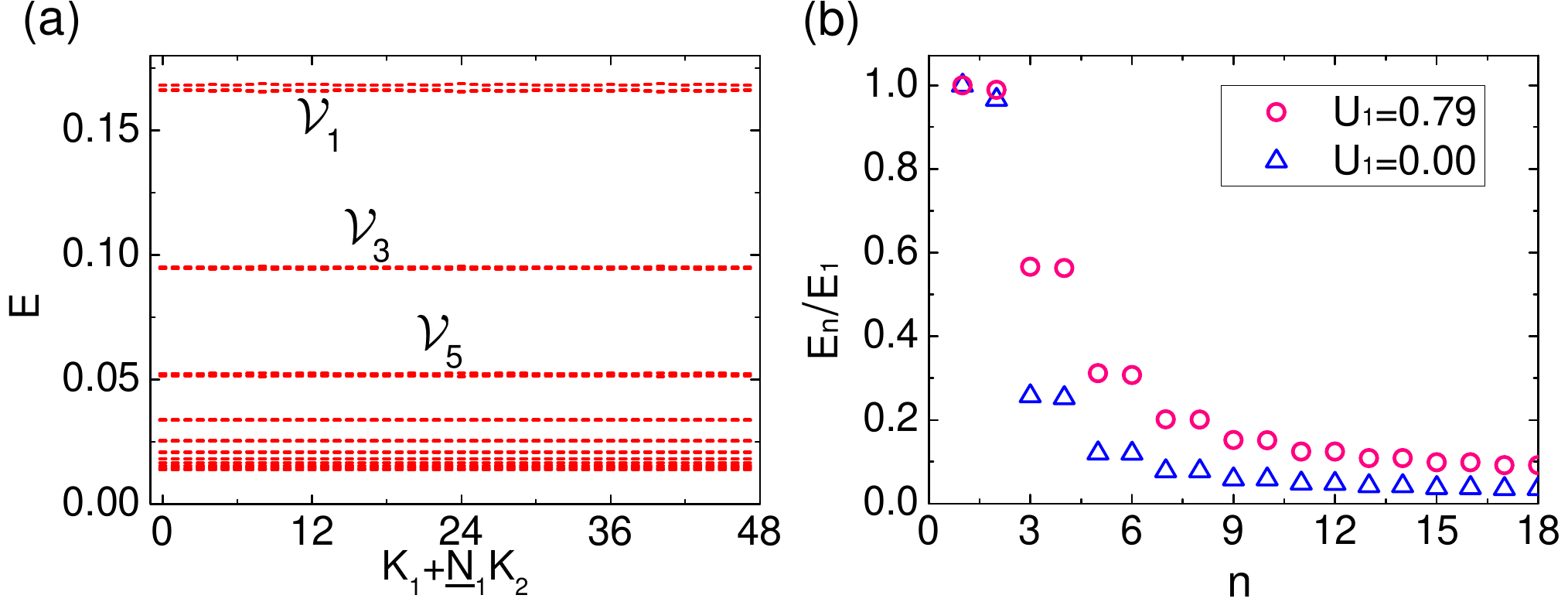}}
\caption{(Color online) The two-particle spectrum analysis of $H_{2\textrm b}$ with $\phi=1/3$, $t'/t=0.16$, $n_{\textrm{max}}=1$ and $U_1=0.79$. (a) The two-particle spectrum on the $N_1\times N_2=12\times12$ lattice. Only the highest levels are plotted. The energy levels form pairs and are identified as Haldane's pseudopotential parameters $\mathcal{V}_1,\mathcal{V}_3,\mathcal{V}_5$. (b) The two-particle energy level $E_n$ (rescaled by the highest level $E_1$) in the $(K_1,K_2)=(0,0)$ sector versus $n$ on the $N_1\times N_2=12\times12$ lattice. The case of pure dipolar interaction ($U_1=0$) is also plotted for comparision.
\label{fg:twoparticle}}
\end{figure}

In Fig.~\ref{fg:twoparticle}(a), we show the high-energy spectrum of two fermions interacting by $H_{2\textrm{b}}$ with the parameters used to stabilize the $\nu=1/2$ MR FCIs. The energy levels form pairs and are almost independent on $(K_1,K_2)$. We can tentatively identify the first (highest) pair as $\mathcal{V}_1$, the second pair as $\mathcal{V}_3$, and the third pair as $\mathcal{V}_5$ (note that the pseudopotential parameters of even order do not appear in the two-particle spectrum for fermions) \cite{andreasprl,PhysRevB.88.205101}. The pairing of energy levels can be easily seen in Fig.~\ref{fg:twoparticle}(b), where we choose $(K_1,K_2)=(0,0)$ sector and rescale the highest energy level as $1$. We find that $\mathcal{V}_3/\mathcal{V}_1\approx0.6$ for $H_{2\textrm{b}}$ with $U_1=0.79$, while $\mathcal{V}_3/\mathcal{V}_1$ is only roughly $0.3$ for the pure dipolar interaction. Considering a large $\mathcal{V}_3/\mathcal{V}_1$ is also crucial for the stabilization of MR states in the second Landau level \cite{PhysRevB.78.155308}, the pseudopotential parameters of our dipolar interaction supplemented by two-body NN attractions are very reasonable.

\begin{figure}
\centerline{
\includegraphics[width=1.0\linewidth]{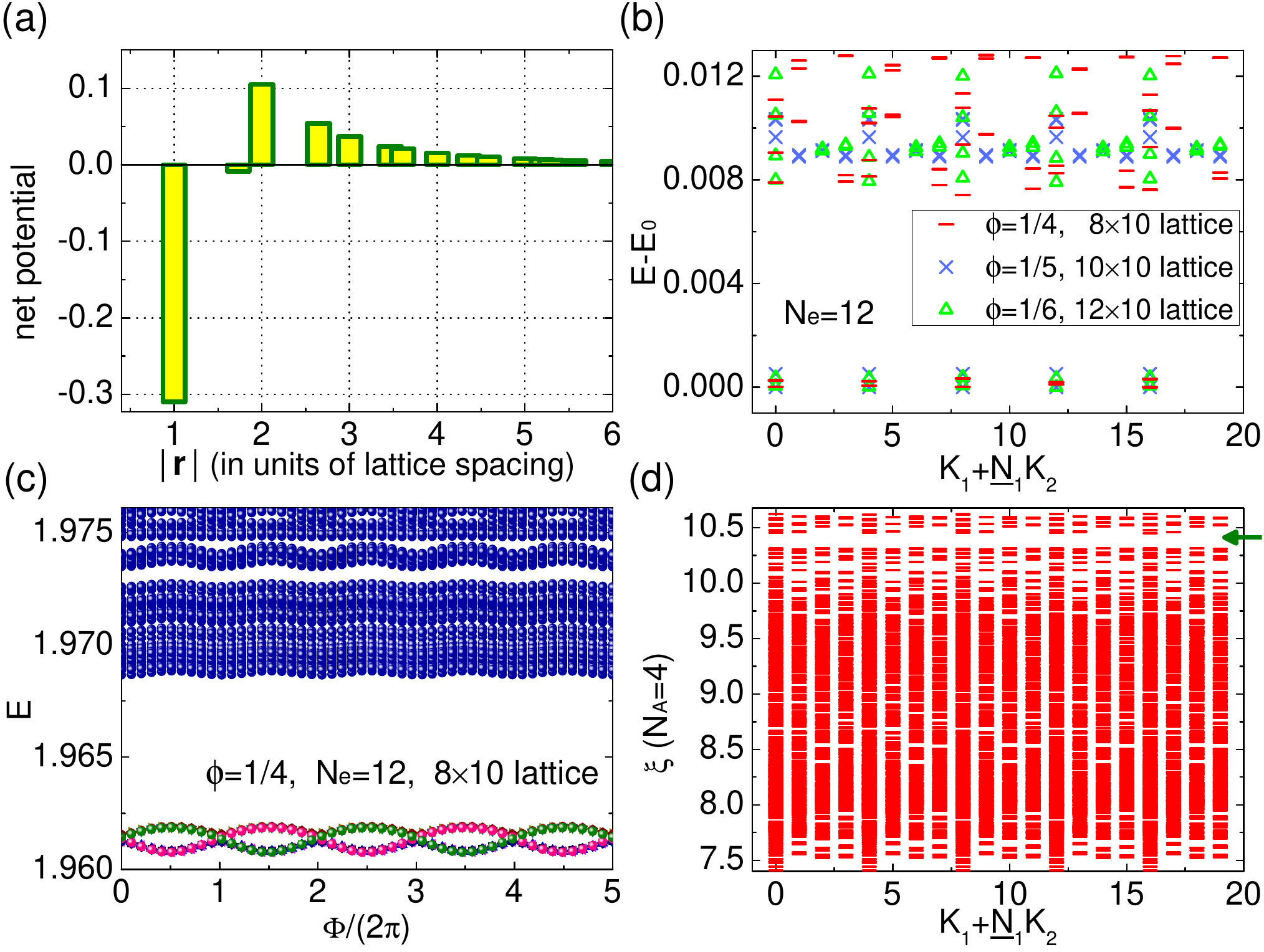}}
\caption{(Color online) Evidence of the $\nu=3/5$ RR FCIs as the ground states of $H_{2\textrm b}$. We choose (~\textbf{\large{--}}~) $n_{\textrm{max}}=1$, $U_1=1.26$, $t'/t=0.11$ for $\phi=1/4$; ({\boldmath\large$\times$}) $n_{\textrm{max}}=3$, $(U_1,U_2,U_3)=(1.31,0.2,0.02)$, $t'/t=0.10$ for $\phi=1/5$; and ({\boldmath$\triangle$}) $n_{\textrm{max}}=1$, $U_1=1.55$, $t'/t=0.08$ for $\phi=1/6$. (a) The net potential for the combination of dipolar interaction and attractive NN, NNN, and NNNN interactions as a function of distance $r$ on the lattice with $n_{\textrm{max}}=3$, $(U_1,U_2,U_3)=(1.31,0.2,0.02)$. The attractions are so strong that net potentials for NN and NNN sites become negative. (b) The low-energy spectrum for $N_{e}=12$ at $\nu=3/5$ with different flux densities $\phi=1/4,1/5$ and $1/6$ on $N_1\times N_2=8\times10$, $10\times10$, $12\times10$ lattices, respectively. (c) The $x$-direction spectral flow for $N_{\textrm e}=12$ with $\phi=1/4$ on the $N_1\times N_2=8\times10$ lattice. (d) The $N_{ A}=4$ PES for $N_{e}=12$ with $\phi=1/6$ on the $N_1\times N_2=12\times10$ lattice. The number of states below the entanglement gap (indicated by the green arrow) is $4765$.
\label{fg:RR_states}}
\end{figure}

\section{$Z_3$ Read-Rezayi FCIs at $\nu=3/5$}
Compared with the $\nu=1/2$ MR FCIs, the $\nu=3/5$ $Z_3$ Read-Rezayi (RR) FCIs are more appealing because the Fibonacci anyon excitations of these states can be used to perform universal quantum computation (the Majorana anyon excitations of the MR FCIs cannot). However, the RR FCIs are more fragile and sensitive to the interactions and sample sizes. In order to stabilize these state, we need finer tuning of the interaction than what we do in the search of MR FCIs.

The higher filling fraction and its odd denominator make the number of available lattice samples accessible by exact diagonalization at $\nu=3/5$ is much less than that at $\nu=1/2$. However, we still obtain encouraging evidence of the RR FCIs at high flux densities. By setting appropriate $n_{\textrm{max}}$ and $U_m$, we observe ten quasidegenerate ground states in the low-energy spectrum for $N_{ e}=12$ fermions with $\phi=1/4,1/5$ and $1/6$ [Fig.~\ref{fg:RR_states}(b)]. This ten-fold degeneracy is robust against the twisted boundary conditions [Fig.~\ref{fg:RR_states}(c)]. The counting of levels below the gap of the PES also matches the requirement of the $(3,5)$-admissible rule \cite{Bernevig_PRB85}[Fig.~\ref{fg:RR_states}(d)]. All these evidences support that the ground states at $\nu=3/5$ are the RR FCIs.

\section{Conclusion} 
In this paper, we have demonstrated that a combination of long-range dipolar interaction and two-body short-range attractions for interacting fermions on a triangular lattice can exhibit ground states as non-Abelian fractional Chern insulators. Our single-particle model is a simple generalization of the triangular Hofstadter model by adding an extra next-nearest-neighbor hopping. This extra term is crucial for tuning the lowest band to be nearly flat. After switching on interactions in this flatband, we have observed robust $\nu=1/2$ Moore-Read FCIs for flux densities as high as $1/3$. Besides the topological degeneracy, spectral flow and entanglement spectrum, the adiabatic continuity to the ground states of the three-body interaction also proves the ground states of our two-body long-range interaction are indeed in the MR phase. We compute the two-fermion energy spectrum and extract the Haldane's pseudopotential parameters of our long-range interaction, which are reasonable compared with the known results in Landau levels. The encouraging evidence is also discovered for the more exotic $\nu=3/5$ $Z_3$ Read-Rezayi FCIs at flux densities as high as $1/4$.
The interactions discussed in our scheme are quite promising to be realized \cite{DipolarExp1,DipolarExp2,DipolarExp3,PhysRevLett.103.080406}.
Considering the recent successful experimental realizations of the Hofstadter model \cite{hofstadterexp1,hofstadterexp2}, our results may provide insights into the experimental preparation of fermionic FCIs in optical lattices.

It is promising that the fermionic non-Abelian FCIs can be similarly stabilized by appropriate two-body long-range interactions in Chern bands of other lattice models \cite{unpublish}, including those with higher Chern number bands \cite{PhysRevB.86.241111,liu2012fractional}. We highlight the physical significance of our scheme for the realistic dipolar interaction, applicable possibly in various lattice configurations and crucial role in topological quantum computation.

Z.~L. thanks E.~J.~Bergholtz and E.~Kapit for related collaborations and N.~Regnault for discussions. Z.~L. was supported by the Department of Energy, Office of Basic Energy Sciences through Grant No.~DE-SC0002140. This work was partially supported by NSFC (11175248).

\bibliography{Triangular}

\end{document}